\documentclass[english,twocolumn,twoside]{IEEEtran}

\usepackage{amsmath}
\usepackage{amssymb}
\usepackage{amscd}
\usepackage{babel}
\usepackage{latexsym}
\usepackage{epsfig}
\usepackage{bbm}

\newtheorem{defi}{Definition}
\newtheorem{lemma}[defi]{Lemma}
\newtheorem{satz}[defi]{Theorem}
\newtheorem{cor}[defi]{Corollary}
\newtheorem{rem}[defi]{Remark}

\newtheorem{conj}[defi]{Conjecture}
\newtheorem{exempel}[defi]{Example}
\newtheorem{quaest}[defi]{Question}

\DeclareRobustCommand\openone{\leavevmode\hbox{\small1\normalsize\kern-.30em1}}%

\newcommand{\qed}{\hfill $\blacksquare$}
\newcommand{\tr}{{\operatorname{Tr}\,}}

\newcommand{\supp}{{\operatorname{supp}\,}}

\newcommand{\R}{{\mathbb{R}}}

\newcommand{\E}{{\operatorname{\mathbb{E}}}}
\newcommand{\alg}[1]{{\mathfrak{#1}}}
\newcommand{\fset}[1]{{\mathcal{#1}}}

\newcommand{\1}{{\openone}}
\newcommand{\markov}{{\,\text{---}\!\!\!\ominus\!\!\!\text{---}\,}}

\newlength{\blank}
\newenvironment{beweis}[1][{\hspace{-\blank}}]{{\noindent\emph{Proof~{#1}.\ }}}{\hfill $\blacksquare$\vskip 0.5\baselineskip}

\begin{document}


\title{Compression of sources of probability\protect\\
 distributions and density operators}
\author{Andreas Winter\thanks{The author is with Department of Computer Science, University of Bristol, Merchant Venturers Building, Woodland Road, Bristol BS8 1UB, U.K.}\thanks{Email: \texttt{winter@cs.bris.ac.uk}}\thanks{Dated 21${}^{\rm th}$ August, 2002.}
}
\date{21 August, 2002}

\maketitle


\begin{abstract}
  We study the problem of efficient compression of a stochastic source
  of probability distributions. It can be viewed as a generalization of
  Shannon's source coding problem. It has relation to the theory of
  common randomness, as well as to channel coding and rate--distortion
  theory: in the first two subjects ``inverses''
  to established coding theorems can be derived, yielding a new approach
  to proving converse theorems, in the third we find a new proof
  of Shannon's rate--distortion theorem.
  \par
  After reviewing the known lower bound for the optimal compression rate, we present
  a number of approaches to achieve it by code constructions.
  Our main results are: a better understanding of the known lower
  bounds on the compression rate by means of a strong version of
  this statement, a review of a construction achieving the lower
  bound by using common randomness which we complement by
  showing the optimal use of the latter within a
  class of protocols. Then we review another approach, not dependent on
  common randomness, to minimizing
  the compression rate, providing some insight into its combinatorial
  structure, and suggesting an algorithm to optimize it.
  \par
  The second part of the paper is concerned with the generalization
  of the problem to quantum information theory: the compression
  of mixed quantum states. Here, after reviewing the known lower
  bound we contribute a strong version of it, and discuss the relation
  of the problem to other issues in quantum information theory.
\end{abstract}


\section{Sources of distributions}
\label{sec:pdsources}
A theorem of Shannon~\cite{shannon} basic to all information theory
describes the optimum compression
of a discrete memoryless source, showing that the minimum achievable rate
is the entropy of the source distribution. The situation is the following:
\par
Let $P$ be a probability distribution on the finite set $\fset{X}$.
We call $(E,D)$ an \emph{$(n,\lambda)$--code} for the
\emph{discrete memoryless source $P$}, if
\begin{equation}
  \label{eq:det:code}
  \begin{array}{c}
    E:\fset{X}^n \longrightarrow \fset{C},  \\
    D:\fset{C}   \longrightarrow \fset{X}^n
  \end{array}
\end{equation}
are stochastic maps, with a finite set $\fset{C}$, such that
\begin{equation}
  \label{eq:det:condition}
  \sum_{x^n\in\fset{X}^n} P^n(x^n)\Pr\{x^n=D(E(x^n))\}\geq 1-\lambda,
\end{equation}
where
$$\Pr\{x^n=D(E(x^n))\}=D(E(x^n))\{x^n\}.$$
Denoting the minimal $|\fset{C}|$ such that an $(n,\lambda)$ code exists,
by $M(n,\lambda)$, Shannon~\cite{shannon} shows that for $\lambda\in(0,1)$
$$\lim_{n\rightarrow\infty} \frac{1}{n}\log M(n,\lambda)=H(P),$$
with the entropy $H(P)=-\sum_x P(x)\log P(x)$ of the distribution.
\par
Motivated by the work~\cite{bcfjs}, and by a construction in~\cite{bennett:et:al}
(in footnote 4), we study here the following modification of this problem:
\par
To each $x\in\fset{X}$ is associated a probability distribution $W_x$ on
the finite set $\fset{Y}$ (thus $W$ is a stochastic map, or channel, form
$\fset{X}$ to $\fset{Y}$). An $(n,\lambda)$--code is now a pair $(E,D)$
of stochastic maps
\begin{equation}
  \label{eq:code}
  \begin{array}{c}
    E:\fset{X}^n \longrightarrow \fset{C},  \\
    D:\fset{C}   \longrightarrow \fset{Y}^n
  \end{array}
\end{equation}
(compare with eq.~(\ref{eq:det:code})),
and instead of condition (\ref{eq:det:condition}) we impose
\begin{equation}
  \label{eq:condition}
  \sum_{x^n\in\fset{X}^n} P^n(x^n)\frac{1}{2}\|W^n_{x^n}-D(E(x^n))\|_1\leq \lambda,
\end{equation}
where $\|\cdot\|_1$ is the $\ell^1$--norm on function on $\fset{Y}^n$:
$\|f\|_1=\sum_{y^n} |f(y^n)|$. Note that for two probability distributions
$P$ and $Q$, $\frac{1}{2}\|P-Q\|_1$ equals their
\emph{total variational distance}
$d_{\rm TV}(P,Q)=\sup_{\fset{A}\subset\fset{Y}^n} |P(\fset{A})-Q(\fset{A})|$
of the two.
We define $M(n,\lambda)$ to be the minimal $|\fset{C}|$ of
an $(n,\lambda)$--code.
\par
Note that for $\fset{Y}=\fset{X}$, and $W_x$ the point--mass $\delta_x$
in $x$, the new notion of $(n,\lambda)$--code coincides with
the previous one.
Notice further, that we allow probabilistic choices in the encoding and decoding.
While it is easy to see that this freedom does not help in Shannon's
problem, it is crucial for the more general form, that we will study in this
paper.
\par
The basic problem of course is to find the optimum rate
$$\Gamma_\lambda(P,W)=\lim_{n\rightarrow\infty} \frac{1}{n}\log M(n,\lambda)$$
of compression (if the limit exists; otherwise $\limsup$ is to be considered),
and especially the behaviour of this function at $\lambda\rightarrow 0$.
\par
For the case $\lambda=0$, i.e. perfect restitution of the distributions
$W_x$, these definitions in principle make sense,
but we don't expect a neat theory to emerge. Instead we
define
$$S(n)=\min H(E(P^{\otimes n})),$$
the minimal entropy of the distribution on $\fset{C}$ induced
by the encoder $E$ (with the idea that blocks of these $n$--blocks
we may data compress to this rate). Obviously $S(n_1+n_2)\leq S(n_1)+S(n_2)$,
so the limit
$$\Gamma(P,W)=\lim_{n\rightarrow\infty} \frac{1}{n}S(n)$$
exists, and is equal to the infimum of the sequence.
To evaluate this quantity is another problem we would like to solve.
\par
The structure of this paper is as follows: first we find lower
bounds (section~\ref{sec:lower}), then discuss upper bounds, preferrably
by constructing codes: in section~\ref{sec:cr:trick} we show how
the lower bound is approached by using the additional resource
of common randomness, in section~\ref{sec:local:fid} we prove achievability of
it under a letterwise fidelity criterion as a consequence of this result,
section~\ref{sec:howards} presents a constructions to
upper bound $\Gamma$ and $\Gamma_\lambda$.
In section~\ref{sec:applications} applications of the results
and conjectures are presented:
first, we make it plausible that the distillation procedure
of~\cite{ahlswede:csiszar:1} is asymptotically reversible,
second we show that Shannon's coding theorem allows an ``inverse''
(at least in situations where unlimited common randomness is around),
third we give a simple proof that feedback does not increase the rate
of a discrete memoryless channel,
and fourth demonstrate, how Shannon's rate--distortion theorem follows
as a corollary.
The compression result (with or without common randomness)
thus reveals a great unifying power in classical information theory.
Finally, in section~\ref{sec:quantum} we discuss extensions
of our results to the case of a source of mixed quantum states:
the present discussion fits into this models as probability
distributions are just commuting mixed state density operators.
\par\medskip
Let us mention here the previous work on the problem: the major
initiating works are~\cite{horodecki} and~\cite{bcfjs}. The latter
introduced the distinction between blind and visible coding,
and between the block-- and letterwise fidelity criterion.
In contrast to the pure state case the four possible combinations
of these conditions seem to lead to rather different answers.
The case of blind coding with either the letter-- or blockwise
fidelity criterion was solved recently
by Koashi and Imoto~\cite{koashi:imoto}.
Otherwise in this paper, we will only address the visible case.
An attempt on the letterwise fidelity case with either blind
or visible encoding was made in~\cite{kramer:savari}. However,
an examination of the approach of this work shows that it does
not fit into any of the the classes of fidelity criteria proposed
by~\cite{bcfjs}:
for a code $(E,D)$ one could either apply the \emph{global}
criterion, which is essentially our eq.~(\ref{eq:condition}),
that is definitely not what is considered in~\cite{kramer:savari},
there being employed rate distortion theory.
\par
Or one could impose that the output $E(D(x^n))$ is good
on the average letterwise (the \emph{local} criterion
of~\cite{bcfjs}):
\begin{equation}
  \label{eq:l:condition}
  \sum_{x^n} P^n(x^n)\left[
              \frac{1}{n}\sum_{k=1}^n d\left(W_{x_k},D(E(x^n))_k\right)
                     \right] \leq \lambda,
\end{equation}
where $D(E(x^n))_k$ denotes the marginal distribution of $D(E(x^n))$
on the $k^{\rm th}$ factor in $\fset{Y}^n$, and $d$ is any distance
measure on probability distributions (that we require only to be
convex in the second variable). For $d(P,Q)=\frac{1}{2}\|P-Q\|_1$
this is implied by eq.~(\ref{eq:condition}).
This, too, is not met in~\cite{kramer:savari}, as there $E$ and $D$ are
constructed as deterministic maps, while to satisfy eq.~(\ref{eq:l:condition})
one needs at least a small amount of randomness.
\par
To achieve this one could base the fidelity condition on looking
at individual letter positions of \emph{source and output simultaneously}:
\begin{equation}
  \label{eq:ll:condition}
  \sum_x P(x)\left[\frac{1}{n}\sum_{k=1}^n
          d\left(W_x,\!\!\sum_{x^n:\,x_k=x}\!\!\frac{P^n(x^n)}{P(x)} D(E(x^n))_k\right)
             \right]\! \leq \lambda.
\end{equation}
Condition~(\ref{eq:l:condition}) being weaker than~(\ref{eq:condition}),
this one is still weaker. However, this, too, does not coincide with
the criterion of~\cite{kramer:savari}: denoting by $G$ the
joint distribution of $x$ and $y$ according to $P$ and $W$,
i.e. $G(xy)=P(x)W_x(y)$, one considers
\begin{equation}
  \label{eq:ks:condition}
  \sum_{x^n} P^n(x^n) d\!\left(G,\frac{1}{n}
                              \sum_{k=1}^n \delta_{x_k}\otimes D(E(x^n))_k
                         \right) \leq \lambda
\end{equation}
(This is implied by eq.~(1) of~\cite{kramer:savari} for
$\epsilon=\delta=\lambda/2$, which in turn  is implied by
eq.~(\ref{eq:ll:condition}) for $\epsilon=\delta=\sqrt{\lambda}$).
It is not at all clear how to connect this
with any of the above: eq.~(\ref{eq:ks:condition}) is about the
\emph{empirical joint distribution} of letters in $x^n$
and $D(E(x^n))$ (assume for simplicity, as indeed the authors
of~\cite{kramer:savari} do, that $E$ and $D$ are deterministic),
that is about a distribution created by selecting a position $k$
randomly, while eqs.~(\ref{eq:condition}) to~(\ref{eq:ll:condition})
are about distributions created either by the coding process
alone or in conjunction with the source. Our view is confirmed
in an independent recent analysis of~\cite{kramer:savari}
by Soljanin~\cite{soljanin}, to the same effect.
\par
An interesting new twist was added when in~\cite{bennett:et:al}
(and later in a more extended way in~\cite{duer:vidal:cirac}
and the recent~\cite{soljanin})
the use of unlimited common randomness between the sender and
receiver was allowed in the visible coding model with blockwise
fidelity criterion. As already mentioned,
we reproduce this result here in detail,
with special attention to the resource of common randomness:
we present a protocol for which we prove that it has minimum
common randomness consumption in the class of protocols
which even simulate full passive feedback of the received signal
to the sender.

\section{Lower bound and conjectures}
\label{sec:lower}
Let the random variable $X^n=(X_1,\ldots,X_n)$ be
distributed according to $P^n$. Then we can define $Y^n$ by
$$\Pr\{Y^n=y^n|X^n=x^n\}=W^n_{x^n}(y^n).$$
By (\ref{eq:condition}) we have the Markov chain
$$X^n\markov E(X^n)\markov D(E(X^n))\approx Y^n.$$
Using data processing inequality as follows:
\begin{equation*}\begin{split}
  \log|\fset{C}| &\geq H(E(X^n)) \\
                 &\geq I(X^n\wedge E(X^n)) \\
                 &\geq I(X^n\wedge D(E(X^n)) \\
                 &\geq I(X^n\wedge Y^n)-n f(\lambda),
\end{split}\end{equation*}
with $f(\lambda)\rightarrow 0$ for $\lambda\rightarrow 0$.
To be precise, one may choose (for $\lambda\leq 1/2$)
$$f(\lambda)=\lambda(\log|\fset{X}|+2\log|\fset{Y}|)+2h(\lambda),$$
employing the following well known result with eq.~(\ref{eq:condition}).
\begin{lemma}
  \label{lemma:H:cont}
  Let $P$ and $Q$ be probability distributions on a set
  with finite cardinality $a$, such that
  $\|P-Q\|_1\leq 1/2$. Then
  $$|H(P)-H(Q)|\leq a\,h\!\left(\frac{\lambda}{a}\right)
               :=    -\lambda\log\frac{\lambda}{a}.$$
\end{lemma}
\begin{beweis}
  See e.g.~\cite{csiszar:koerner}.
\end{beweis}
Thus we arrive at
\begin{satz}
  \label{satz:lower}
  For any $n$ and $0<\lambda<1$:
  $$\frac{1}{n}\log M(n,\lambda)\geq I(P;W)-f(\lambda),$$
  where
  $$I(P;W)=H(PW)-\sum_x P(x)H(W_x)$$
  is the mutual information of the channel $W$ between the
  input distribution $P$ and the output
  distribution $PW=\sum_x P(x)W_x$.
  \qed
\end{satz}
By using slightly stronger estimates, we even get
\begin{satz}
  \label{satz:strong:lower}
  For every $\lambda\in(0,1)$
  $$\liminf_{n\rightarrow\infty}\frac{1}{n}\log M(n,\lambda)\geq I(P;W).$$ 
\end{satz}
\begin{beweis}
  Let $(E,D)$ be an optimal $(n,\lambda)$--code. From eq.~(\ref{eq:condition})
  we find (by a Markov inequality argument) that
  $$P^n\left\{x^n:\frac{1}{2}\|W^n_{x^n}-D(E(x^n))\|_1\leq\sqrt{\lambda}
       \right\} \geq 1-\sqrt{\lambda}.$$
  Denote the intersection of this set with the typical
  sequences $\fset{T}^n_{P,\delta}$ (see eq.~(\ref{eq:typical})
  below) by $\fset{A}$, with
  $\delta=\sqrt{\frac{2|\fset{X}|}{1-\sqrt{\lambda}}}$. Then
  $$P^n(\fset{A})\geq \frac{1-\sqrt{\lambda}}{2}=:\lambda',$$
  and there exists an $(n,\lambda')$--transmission code
  $\fset{U}\subset\fset{A}$ for the channel $W^n$ with 
  $|\fset{U}|\geq \exp(nI(P;W)-O(\sqrt{n}))$, see~\cite{csiszar:koerner}
  (the case of a classical--quantum channel $W$ was done
  in~\cite{winter:diss}). By construction this is a
  $(n,1-\lambda')$--code
  for the channel $D\circ E$.
  \par
  We want now view $E$ as belonging to the message encoder,
  and $D$ as belonging to the message decoder, the resulting code
  being one for the identical channel on $\fset{C}$.
  Let us denote the concatenation of the map $D$ with the
  channel decoder by $\delta$. On the other hand, we may
  replace $E$ by a deterministic map $\varepsilon$,
  because randomization at the encoder never decreases
  error probabilities: $(\varepsilon,\delta)$ still is
  an $(n,1-\lambda')$--code. It is now obvious that
  $|\varepsilon^{-1}(c)|\leq \lambda^{\prime -1}$
  for every $c\in\fset{C}$, hence
  $$M(n,\lambda)=|\fset{C}|\geq \lambda'|\fset{U}|
                           =    \exp\left(nI(P;W)-O(\sqrt{n})\right),$$
  and we are done.
\end{beweis}
It might be a bit daring to formulate conjectures at this point, so
we content ourselves with posing the following questions:
\begin{quaest}
  \label{quest:main}
  Is it true that for all $\lambda\in(0,1)$
  $$\lim_{n\rightarrow\infty}\frac{1}{n}\log M(n,\lambda)= I(P;W)\ ?$$
\end{quaest}
In fact, we would like to go present a slightly stronger statement:
\par
\emph{Question~\ref{quest:main}'}: For every $\lambda\in(0,1)$, $\epsilon>0$, $\delta>0$,
  and large enough $n$ does there exists a $(n,\lambda)$--code with
  $$\frac{1}{n}\log|\fset{C}|\leq I(P;W)+\epsilon$$
  and with the additional property that
  $$\forall x^n\in\fset{T}^n_{P,\delta}\quad
                          \frac{1}{2}\|W^n_{x^n}-D(E(x^n))\|_1\leq \lambda\ ?$$
\par
Here $\fset{T}^n_{P,\delta}$ is the set of \emph{typical sequences}:
\begin{equation}
  \label{eq:typical}
  \fset{T}^n_{P,\delta}=\left\{
           x^n:\ \forall x\ |N(x|x^n)-nP(x)|\leq \delta\sqrt{n}\sigma_x
                          \right\},
\end{equation}
where $N(x|x^n)$ counts the number of occurences of $x$ in $x^n$,
and $\sigma_x:=\sqrt{P(x)(1-P(x))}$. Observe that by Chebyshev's
inequality
\begin{equation}
  \label{eq:typical:prob}
  P^n\left(\fset{T}^n_{P,\delta}\right)\geq 1-\frac{|\fset{X}|}{\delta^2}.
\end{equation}
In fact, by employing the Chernoff bound we even obtain
\begin{equation}
  \label{eq:typical:prob:exp}
  P^n\left(\fset{T}^n_{P,\delta}\right)\geq 1-|\fset{X}|\exp(-\delta^2).
\end{equation}
With these bounds it is easily seen that a positive answer to the latter
question implies the same to the former. But also conversely, it is not difficult
to show that a ``yes'' to question~\ref{quest:main} implies a ``yes''
to question~\ref{quest:main}'.

\section{... and how to achieve it (cheating slightly)}
\label{sec:cr:trick}
The following construction is a generalization and refinement of the
one by Bennett et~al.~\cite{bennett:et:al} (footnote 4), found
independently by D\"ur, Vidal, and Cirac~\cite{duer:vidal:cirac}.
The idea there is to
use common randomness between the sender and the receiver of the
encoded messages. Formally this means that $E$ and $D$ also depend on
a common random variable $\nu$,
uniformly distributed and independent of all others.
Note that this has a nice expression when viewing $E$ and $D$ as map
valued random variables: here we allow dependence (via $\nu$) between $E$
and $D$, while in the initial definition, eq.~(\ref{eq:code}), $E$ and $D$
are independent (as random variables). It seems that the power of allowing
the use of common randomness can be understood from this point of view:
it is a ``convexification'' of the theory with deterministic or independent
encoders and decoders.
\par
It is easy to see that the lower bound of theorem~\ref{satz:lower}
still applies here. We only have to modify the derivation a
little bit:
\begin{equation*}\begin{split}
  \log|\fset{C}| &\geq H(E(X^n)|\nu) \\
                 &\geq I(X^n\wedge E(X^n)|\nu) \\
                 &\geq I(X^n\wedge D(E(X^n)|\nu) \\
                 &\geq I(X^n\wedge Y^n|\nu)-n \tilde{f}(\lambda) \\
                 &=    I(X^n\wedge Y^n)-n \tilde{f}(\lambda),
\end{split}\end{equation*}
with a slight variant $\tilde{f}$ of $f$.
\par\medskip
We shall apply an explicit large deviation estimate for
sampling probability distributions from~\cite{ahlswede:performance}
(extended to density operators in~\cite{ahlswede:winter:QID}),
which we state separately without proof:
\begin{lemma}
  \label{lemma:large:deviation}
  Let $X_1,\ldots,X_M$ be independent identically distributed
  (i.i.d.) random variables with
  values in the function algebra on
  the finite set ${\cal K}$, which are bounded between
  $0$ and $\1$, the constant function with value $1$.
  Assume that the average $\E X_\mu=\sigma\geq s\1$.
  Then for $0<\eta<1/2$
  $$\Pr\left\{\frac{1}{M}\sum_{\mu=1}^M X_\mu\not\in[(1\pm\eta)\sigma]\right\}
                 \leq 2|\fset{K}|\exp\left(-M\frac{\eta^2 s}{2\ln 2}\right)\!,$$
  where $[(1\pm\eta)\sigma]=[(1-\eta)\sigma;(1+\eta)\sigma]$ is an interval
  in the value--wise order of functions:
  $[A;B]=\{X:\forall k\ A(k)\leq X(k)\leq B(k)\}$.
  \qed
\end{lemma}
Before we prove our main theorem, we need three lemmas on exact
types and conditional types. The first is a simple yet crucial
observation:
\begin{lemma}
  \label{lemma:types:blurb}
  Let $W$ be a channel from $\fset{X}$ to $\fset{Y}$, $P$ a p.d.
  on $\fset{X}$, $Q=PW$ the induced distribution on $\fset{Y}$
  and $V$ the transpose channel from $\fset{Y}$ to $\fset{X}$.
  \par
  Let $R$, $S$ be exact $n$--types of $\fset{X}$, $\fset{Y}$,
  respectively that are marginals of a joint exact
  $n$--type $T$ of $\fset{X}\times\fset{Y}$.
  Consider the uniform distribution
  $P^n_R$ on $\fset{T}^n_R$ on $\fset{T}^n_R$, which has the property
  $$P^n_R(x^n)=\frac{1}{|\fset{T}^n_R|}=\frac{P^n(x^n)}{P^n(\fset{T}^n_R)}
                                          \quad(\text{for }x^n\in\fset{T}^n_R),$$
  and the channel from $\fset{T}^n_R$ to $\fset{T}^n_S$,
  \begin{equation*}\begin{split}
    W^n_T(y^n|x^n) &=\frac{1}{|\fset{T}^n_T(x^n)|}1_T(x^ny^n)         \\
                   &=\frac{|\fset{T}^n_R|}{|\fset{T}^n_T|}
                     =\frac{W^n(y^n|x^n)}{W^n(\fset{T}^n_T(x^n)|x^n)}
                       \quad(\text{for }x^ny^n\in\fset{T}^n_T),
  \end{split}\end{equation*}
  where $\fset{T}^n_T(x^n):=\fset{T}^n_T\cap\left(\{x^n\}\times\fset{T}^n_S\right)$
  is the set of \emph{conditional exact typical sequences} of $x^n$.
  \par
  Then the induced distribution $Q^n_S=P^n_R W^n_{T}$
  on $\fset{T}^n_S$ is the uniform distribution, i.e.
  $$Q^n_S(y^n)=\frac{1}{|\fset{T}^n_S|}=\frac{Q^n(y^n)}{Q^n(\fset{T}^n_S)}
                                          \quad(\text{for }y^n\in\fset{T}^n_S),$$
  and the transpose channel to $W^n_{T}$ is indeed $V^n_{T}$,
  defined by
  \begin{equation*}\begin{split}
    V^n_T(x^n|y^n) &=\frac{1}{|\fset{T}^n_T(y^n)|}1_T(x^ny^n)         \\
                   &=\frac{|\fset{T}^n_S|}{|\fset{T}^n_T|}
                     =\frac{V^n(x^n|y^n)}{V^n(\fset{T}^n_T(y^n)|y^n)}
                       \quad(\text{for }x^ny^n\in\fset{T}^n_T),
  \end{split}\end{equation*}
  with $\fset{T}^n_T(y^n):=\fset{T}^n_T\cap\left(\fset{T}^n_R\times\{y^n\}\right)$.
\end{lemma}
\begin{beweis}
  Straightforward.
\end{beweis}
\begin{lemma}
  \label{lemma:cardinalities}
  There is an absolute constant $K$ such that for all distributions $P$ on $\fset{X}$,
  $x^n\in\fset{T}^n_R$, channels $W:\fset{X}\rightarrow\fset{Y}$ and $\delta>0$
  \begin{align*}
    |\fset{T}^n_{P,\delta}|      &\leq \exp\bigl(nH(P)+K\delta|\fset{X}|\sqrt{n}\bigr),           \\
    |\fset{T}^n_{P,\delta}|      &\geq \exp\bigl(nH(P)-K\delta|\fset{X}|\sqrt{n}\bigr),                \\
    |\fset{T}^n_{W,\delta}(x^n)| &\leq \exp\bigl(nH(W|R)+K\delta|\fset{X}\!\times\!\fset{Y}|\sqrt{n}\bigr), \\
    |\fset{T}^n_{W,\delta}(x^n)| &\geq \exp\bigl(nH(W|R)-K\delta|\fset{X}\!\times\!\fset{Y}|\sqrt{n}\bigr).
  \end{align*}
  For $\delta=0$, consider a joint $n$--type $T$ on $\fset{X}\times\fset{Y}$ with
  marginals $R$ on $\fset{X}$ and $S$ of $\fset{Y}$. Then, introducing the
  channel $Z$ with $T(xy)=R(x)Z(y|x)$:
  \begin{align*}
    |\fset{T}^n_R|      &\leq \exp(nH(R)),                       \\
    |\fset{T}^n_R|      &\geq (n+1)^{-|\fset{X}|}\exp(nH(R)),        \\
    |\fset{T}^n_T(x^n)| &\leq \exp(nH(Z|R)),                             \\
    |\fset{T}^n_T(x^n)| &\geq (n+1)^{-|\fset{X}\times\fset{Y}|}\exp(nH(Z|R)).
  \end{align*}
\end{lemma}
\begin{beweis}
  See~\cite{wolfowitz}.
\end{beweis}
The third contains the central insight for our construction:
\begin{lemma}
  \label{lemma:covering}
  With the hypotheses and notation of lemma~\ref{lemma:types:blurb}
  there exist families $(Y^{(\nu)}_\mu)_{\mu=1,\ldots,M}$,
  $\nu=1,\ldots,N$, from $\fset{T}^n_{S}$ such that for all $\nu$
  \begin{equation}
    \frac{1}{M}\sum_\mu V^n_T(\cdot|Y^{(\nu)}_\mu)
            \in\left[(1-\epsilon)P^n_R,(1+\epsilon)P^n_R\right],\tag{I${}_\nu$}
  \end{equation}
  and
  \begin{equation}
    \frac{1}{NM}\sum_{\nu\mu} \delta_{Y^{(\nu)}_\mu}
            \in\left[(1-\epsilon)Q^n_S,(1+\epsilon)Q^n_S\right],\tag{II}
  \end{equation}
  for all $M$ and $N$ that satisfy
  \begin{align*}
    M  &> \frac{2\ln 2}{\epsilon^2}
            \frac{|\fset{T}^n_R|\,|\fset{T}^n_S|}{|\fset{T}^n_T|}
              \log\left(4N|\fset{T}^n_R|\right),                  \\
    NM &> \frac{2\ln 2}{\epsilon^2}
            |\fset{T}^n_S|\log\left(4|\fset{T}^n_S|\right).
  \end{align*}
\end{lemma}
\begin{beweis}
  Introduce i.i.d.~random variables, distributed on $\fset{T}^n_S$
  according to $Q^n_S$ (i.e. uniformly). Then for all $\nu.\mu$:
  $$\E\delta_{Y^{(\nu)}_\mu}=Q^n_S,\quad
                    \E V^n_T(\cdot|Y^{(\nu)}_\mu)=P^n_R.$$
  Hence lemma~\ref{lemma:large:deviation} applies and we find
  \begin{align*}
    \forall\nu\ &\Pr\{\neg\text{I}_\nu\} \leq
                  2|\fset{T}^n_R|\exp\left(
                       -M\frac{\epsilon^2|\fset{T}^n_T(y^n)|}{2\ln 2|\fset{T}^n_R|}
                                     \right), \\
    \text{and}\ &\Pr\{\neg\text{II}\} \leq
                  2|\fset{T}^n_S|\exp\left(
                       -NM\frac{\epsilon^2}{2\ln 2|\fset{T}^n_S|}\right).
  \end{align*}
  By choosing $N$ and $M$ according to the lemma we enforce that
  the sum of these probabilities is less than $1$, hence there
  are actual values of the $Y^{(\nu)}_\mu$ such that
  all (I${}_\nu$) and (II) are satisfied.
\end{beweis}
With this we are ready to prove:
\begin{satz}
  \label{satz:sim:feedback:channel}
  There exists an $(n,\lambda)$--code
  $(E_\nu,D_\nu)_{\nu=1\ldots N}$ with
  $$|\fset{C}|\leq \exp(nI(P;W)+O(\sqrt{n}))$$
  and common randomness consumption
  $$N\leq \exp(nH(W|P)+O(\sqrt{n})).$$
  In fact, not only the condition (\ref{eq:condition}) is satisfied but
  the even stronger
  \begin{equation}
    \label{eq:individual:condition}
    \forall x^n\in\fset{T}^n_{P,\delta}\quad 
                          \frac{1}{2}\|W^n_{x^n}-D(E(x^n))\|_1\leq \lambda.
  \end{equation}
\end{satz}
\begin{beweis}
  Suppose $x^n$ is seen at the source, and that its type is $R$.
  For each joint $n$--type $T$ of $\fset{X}\times\fset{Y}$
  we assume that families $(Y^{(\nu)}_\mu)$ as
  described in lemma~\ref{lemma:covering}
  are fixed throughout.
  \par
  Then the protocol the sender follows is:
  \begin{enumerate}
    \item Choose a joint type $T$ on $\fset{X}\times\fset{Y}$ with probability
      $W^n(\fset{T}^n_{T}(x^n)|x^n)$ and send it. Note that $T$ can be written
      $T(xy)=R(x)Z(y|x)$, with the marginal $R$ on $\fset{X}$ and a channel
      $Z:\fset{X}\rightarrow\fset{Y}$.
    \item If $R$ is not typical or $T$ is not jointly typical then terminate.
    \item Use the common randomness to choose $\nu$ uniformly.
    \item Choose $\mu$ according to
      $$\Pr\{\mu|x^n\}=\frac{W^n_T(Y^{(\nu)}_\mu|x^n)}{\sum_{\mu'} W^n_T(Y^{(\nu)}_{\mu'}|x^n)},$$
      and send it.
  \end{enumerate}
  \par
  The receiver chooses $y^n=Y^{(\nu)}_\mu$, using the common randomness
  sample $\nu$.
  Let us first check that this procedure works correctly:
  \par
  For typical $x^n$ we can calulate the distribution of $y^n$ conditional on the
  event that their joint type is $T$: this is then a distribution on $\fset{T}^n_T(x^n)$,
  and we assume $T$ to be typical.
  \begin{equation*}\begin{split}
    \Pr\{\cdot|x^n,T\}
                &=\frac{1}{N}\sum_{\nu,\mu=1}^{N,M}
                               \frac{W^n_T(Y^{(\nu)}_\mu|x^n)}{\sum_{\mu'} W^n_T(Y^{(\nu)}_{\mu'}|x^n)}
                               \delta_{Y^{(\nu)}_\mu}                                                  \\
                &=\frac{1}{NM}\sum_{\nu,\mu=1}^{N,M}
                                \frac{P^n_R(x^n)}{Q^n_S(y^n)}
                                \frac{W^n(y^n|x^n)}{\frac{1}{M}\sum_{\mu'} V^n_T(x^n|Y^{(\nu)}_{\mu'})}
                                \delta_{Y^{(\nu)}_\mu}                                                 \\
                &=\frac{1}{NM}\sum_{\nu,\mu=1}^{N,M}
                                \frac{1}{1+B(\epsilon)}\frac{W^n(y^n|x^n)}{Q^n_S(y^n)}
                                \delta_{Y^{(\nu)}_\mu}                                                 \\
                &=\frac{1+B(\epsilon)}{1+B(\epsilon)} W^n(y^n|x^n),
  \end{split}\end{equation*}
  with the ``big--B'' notation: $B(\epsilon)$ signifies any function
  whose modulus is bounded by $\epsilon$.
  Here we have used the definition of the protocol, then lemma~\ref{lemma:types:blurb}
  (for the definition of $V^n_T$ and the fact that $W^n_T(y^n|x^n)$ does not depend on
  $y^n\in\fset{T}^n_T(x^n)$), then lemma~\ref{lemma:covering}. So, the induced
  distribution is, up to a factor between $\frac{1-\epsilon}{1+\epsilon}$
  and $\frac{1+\epsilon}{1-\epsilon}$, equal to the correct output distribution $W^n_T(\cdot|x^n)$.
  Now averaging over the typical $T$ gives eq.~(\ref{eq:individual:condition}).
  \par
  What is the communication cost? Sending $T$ is asymptotically for
  free, as the number of joint types is bounded by the polynomial
  $(n+1)^{|\fset{X}\times\fset{Y}|}$. Sending $\mu$ costs $\log M$
  bits, with $M$ bounded according to lemma~\ref{lemma:covering}.
  That is,
  \begin{equation*}\begin{split}
    \log M &\leq n\left(\max_{T\text{ typical}} I(R;Z)\right)+O(\log n) \\
           &\leq nI(P;W)+O(\sqrt{n}).
  \end{split}\end{equation*}
  On the other hand
  \begin{equation*}\begin{split}
    \log N &\leq n\left(\max_{T\text{ typical}} \bigl(H(PZ)-I(R;Z)\bigr)\right) \\
           &\leq nH(W|P)+O(\sqrt{n}),
  \end{split}\end{equation*}
  and we are done.
\end{beweis}
\begin{rem}
  \label{rem:chernoff:better}
  In the above statement of theorem~\ref{satz:sim:feedback:channel} we assumed $\lambda$
  to be a constant, absorbed into the ``$O(\sqrt{n})$'' in the
  code length estimate. Using the Chernoff estimate~(\ref{eq:typical:prob:exp})
  on the probabilities of typical sets in the above proof in fact
  shows the existence of an $(n,\lambda)$--code
  satisfying~(\ref{eq:individual:condition})
  $$|\fset{C}|\leq\exp\bigl(nI(P;W)+O(-\log\lambda)\sqrt{n}\bigr).$$
\end{rem}
\par
In the line of~\cite{bennett:et:al}, the interpretation of this
result is that investing common randomness at rate $H(W|P)$,
one can simultate the noisy channel $W$ by a noiseless one
of rate $I(P;W)$, when sending only $P$--typical words.
\par
Considering the construction again, we observe that in fact
not only it provides a simulation of the channel $W$, but
additionally of the \emph{noiseless passive feedback}. Simply because
the sender can read off from his random choices the $y^n$ obtained
by the receiver, too. This observation is the key to show
that our above construction is optimal under the hypothesis
that the channel \emph{with noiseless passive feedback} is simulated:
in fact, since both sender and receiver can observe the
very output sequence $y^n$ of the channel, which has entropy
$H(PW)$, they are able to generate common randomness at this rate.
Since communication was only at rate $I(P;W)$, the difference
must by invested in prepared common randomness: otherwise we
would get more of it out of the system than we could have possibly
invested. Formally this insight is captured by the following result:
\begin{satz}
  \label{satz:feedback:channel:opt}
  If the decoder of a $(n,\lambda)$--code $(E,D)$ with common
  randomness consumption $\nu\in[N]$ (with distribution $\xi$)
  depends deterministically
  on $\nu$ and $c\in\fset{C}$ (which is precisely the condition
  that the encoder can recover the receiver's output) then
  \begin{align*}
     |\fset{C}| &\geq \exp\left(n I(P;W)-O(\sqrt{n})\right), \\
    N|\fset{C}| &\geq \exp\left(n H(PW)-O(\sqrt{n})\right).
  \end{align*}
\end{satz}
\begin{beweis}
  For the first inequality introduce the channels
  $A^{(\nu)}_{x^n}=D_\nu(E_\nu(x^n))$, and their induced
  distributions $R^{(\nu)}$ on $\fset{Y}^n$ and transpose
  channels $B^{(\nu)}_{y^n}$ with respect to $P^n$, i.e.
  $$P^n(x^n)A^{(\nu)}_{x^n}(y^n)=R^{(\nu)}(y^n)B^{(\nu)}_{y^n}(x^n).$$
  Then we can rewrite eq.~(\ref{eq:condition}) as
  $$\sum_{y^n} \frac{1}{2}\|Q^n(y^n)V^n_{y^n}
                            -\sum_\nu \xi_\nu R^{(\nu)}(y^n)B^{(\nu)}_{y^n}\|_1
                                                                     \leq \lambda.$$
  This inequality oviously remains valid if we restrict the sum
  to $y^n\in\fset{T}^n_{Q,\delta}$ and replace $V^n_{y^n}$
  and $B^{(\nu)}_{y^n}$ by their restrictions
  to $\fset{T}^n_{V,\delta}(y^n)$: $V^{n\prime}_{y^n}$
  and $B^{(\nu)\prime}_{y^n}$, respectively.
  \par
  On the other hand, choosing
  $\delta=\sqrt{\frac{4|\fset{X}|\,|\fset{Y}|}{1-\lambda}}$,
  we have
  $$\sum_{y^n\in\fset{T}^n_{Q,\delta}} Q^n(y^n)V^{n\prime}_{y^n}(\fset{X}^n)
                 \geq \frac{1+\lambda}{2}=:\lambda',$$
  which yield
  $$\sum_\nu \xi_nu \sum_{y^n\in\fset{T}^n_{Q,\delta}}
          R^{(\nu)}(y^n)B^{(\nu)\prime}_{y^n}(\fset{X}^n) \geq 1-\lambda'.$$
  Hence there exists at least one $\nu$ such that
  $$\sum_{y^n\in\fset{T}^n_{Q,\delta}}
          R^{(\nu)}(y^n)B^{(\nu)\prime}_{y^n}(\fset{X}^n) \geq 1-\lambda'.$$
  Note that, as functions on $\fset{X}^n$,
  $$\sum_{y^n\in\fset{T}^n_{Q,\delta}}
          R^{(\nu)}(y^n)B^{(\nu)\prime}_{y^n}\leq P^n,$$
  so, when we introduce the support $\fset{S}$ of the left hand
  side, we arrive at
  $$P^n(\fset{S})\geq 1-\lambda',$$
  from which our claim follows by a standard
  trick~\cite{wolfowitz}:
  let $\fset{S}'=\fset{S}\cap\fset{T}^n_{P,\delta'}$, with
  $\delta'=\sqrt{\frac{2|\fset{X}|}{1-\lambda'}}$. Then
  $$P^n(\fset{S}')\geq \frac{1-\lambda'}{2},$$
  and using the fact that
  $$\forall x^n\in\fset{T}^n_{P,\delta'}\ 
                 P^n(x^n)\leq\exp(-nH(P)+K|\fset{X}|\delta'\sqrt{n}),$$
  this implies
  $$|\fset{S}|\geq |\fset{S}'|
              \geq \frac{1-\lambda'}{2}\exp(nH(P)-K|\fset{X}|\delta'\sqrt{n}).$$
  Now only note that (since $D_\nu$ is deterministic)
  $$|\fset{S}|\leq |\fset{C}|\max_{y^n\in\fset{T}^n_{Q,\delta}}|
                                            \fset{T}^n_{V,\delta}(y^n)|
              \leq |\fset{C}|\exp(nH(V|Q)+O(\sqrt{n})),$$
  and by $I(P;W)=I(Q;V)=H(P)-H(V|Q)$ we are done.
  \par
  Now for the second inequality:
  from the definition we get, by summing over $x^n$,
  $$\frac{1}{2}\left\|(PW)^n
                 -\sum_\nu \xi_\nu \sum_{x^n} P^n_{x^n}D_\nu(E_\nu(x^n))\right\|_1
                        \leq\lambda.$$
  Because the $D_\nu$ are all deterministic, the distributions
  $D_\nu(E_\nu(x^n))$ are all supported on sets of cardinality
  $|\fset{C}|$. Hence the support $\fset{S}$ of
  $\sum_\nu x_\nu \sum_{x^n} P^n_{x^n}D_\nu(E_\nu(x^n))$
  can be estimated $|\fset{S}|\leq N|\fset{C}|$.
  \par
  On the other hand, we deduce
  $$(PW)^n(\fset{S})\geq 1-\lambda,$$
  which, by the same standard trick~\cite{wolfowitz}
  as before, yields our estimate:
  with $\delta=\sqrt{\frac{2|\fset{Y}|}{1-\lambda}}$,
  the set $\fset{S}'=\fset{S}\cap\fset{T}^n_{PW,\delta}$ satisfies
  $$(PW)^n(\fset{S}')\geq \frac{1-\lambda}{2},$$
  but since for all $y^n\in\fset{T}^n_{PW,\delta}$
  $$(PW)^n(y^n)\leq \exp(-nH(PW)+K|\fset{Y}|\delta\sqrt{n}),$$
  we can conlude
  $$N|\fset{C}|\geq |\fset{S}|\geq |\fset{S}'|
               \geq \frac{1-\lambda}{2}\exp(-nH(PW)+K|\fset{Y}|\delta\sqrt{n}).$$
\end{beweis}
Collecting these results we can state
\begin{cor}
  \label{cor:feedback:channel:opt}
  For any simulation of the channel $W$ together with its
  noiseless passive feedback with error $\lambda<1$, at rate
  $R$ and common randomness consumption rate $C$:
  $$R\geq C(W)=\max_P I(P;W),\quad R+C\geq \max_P H(PW).$$
  Conversely, these rates are also achievable.
\end{cor}
\begin{beweis}
  A simulation of the channel must be in the error
  bound for \emph{every} input $x^n$, hence eq.~(\ref{eq:condition})
  will be satisfied for every distribution $P$.
  The lower bounds follow now from theorem~\ref{satz:feedback:channel:opt}
  by choosing $P$ to maximize $I(P;W)$ and $H(PW)$,
  respectively.
  \par
  To achieve this, the encoder, on seeing $x^n$ reports its
  type to the receiver (asymptotically free) and then they use
  the protocol of theorem~\ref{satz:sim:feedback:channel}
  for $P=P_{x^n}$, the empirical distribution of $x^n$.
  Possibly they have to use the channel at rate $C(W)-I(P;W)$
  to set up additional common randomness beyond
  the given $\max_P H(PW)-C(W)$.
\end{beweis}
\par\medskip
At this point we would like to point out a remarkable parallel
of methods and results to the work~\cite{data-in-qm}: our use of
lemma~\ref{lemma:covering} is the classical case of of the use
of its quantum version from~\cite{ahlswede:winter:QID}, and the main
result of the cited paper is the quantum analog of the
present theorem~\ref{satz:sim:feedback:channel}. The optimality
result there has its classical case formulated in theorems~\ref{satz:lower}
(and~\ref{satz:strong:lower}) and~\ref{satz:feedback:channel:opt},
and even the construction of the following section has its
counterpart there.
\par\medskip
The use of common randomness turned out to be remarkably powerful,
and it is known in various occasions to make problems more tractable: a major
example is the arbitrarily varying channel (see for example the
review~\cite{lapidoth:narayan}).
While for discrete memoryless channels it does not lead to improved rates
or error bounds, it there allows for a ``reverse'' of Shannon's coding
theorem~\cite{bennett:et:al} in the sense of simulating efficiently a noisy
channel by a noiseless one. This viewpoint seems to extend to quantum
channels as well, assisted by entanglement rather than common randomness:
see~\cite{bennett:et:al}. We shall expand on the power of the ``randomness
assisted'' viewpoint in section~\ref{sec:applications}.

\section{Solution under a letterwise criterion}
\label{sec:local:fid}
Here we show that from the theorem of the previous section
a solution to the compression problem under a slightly relaxed
distance criterion follows: whereas previously we had to
employ common randomness to achieve the lower bound
$I(P;W)$, this will turn out to be unnecessary now.
\par
Specifically, our condition will be eq.~(\ref{eq:l:condition}):
\begin{satz}
  \label{satz:coding:local}
  There exists an $n$--block code $(E,D)$ with
  $$|\fset{C}|\leq \exp\left(nI(P;W)+O(\sqrt{n})\right),$$
  such that
  $$\sum_{x^n} P^n(x^n)\frac{1}{n}\sum_{k=1}^n
                          \frac{1}{2}\|W_{x_k}-D(E(x^n))_k\|_1 \leq \lambda.$$
\end{satz}
\begin{beweis}
  Choose an $(n,\lambda)$--code $(E_\nu,D_\nu)_{\nu=1,\ldots,N}$
  as in theorem~\ref{satz:sim:feedback:channel}.
  Obviously this code meets the condition of the theorem,
  except for the use of common randomness. We will show that
  a uniformly random choice among a \emph{small}
  (subexponential) number of $\nu$ is sufficient for this
  to hold. Then the protocol simply is:
  \begin{enumerate}
    \item The sender choses $\nu$ uniformly random (among the
      chosen few), and sends it to the receiver (at asymptotic rate $0$).
    \item She uses $E_\nu$ to encode, and the receiver uses $D_\nu$ to decode.
  \end{enumerate}
  By construction this meets the requirements of the theorem.
  \par
  To prove our claim, note that from theorem~\ref{satz:sim:feedback:channel}
  we can infer
  $$\forall x^n\in\fset{T}^n_{P,\delta} \forall k\quad
      \frac{1}{2}\left\|W_{x_k}-\sum_\nu \xi_\nu D_\nu(E_\nu(x^n))_k\right\|_1
                                                              \leq \epsilon.$$
  Introduce i.i.d.~random variables $T_1,\ldots,T_Q$, distributed
  according to $\xi_\nu$. With the notations
  $X^{(\nu)}_{x^n}=D_\nu(E_\nu(x^n))$ and
  $X^{(\nu)}_{x^n|k}=D_\nu(E_\nu(x^n))_k$ we have
  \begin{align*}
    \E X^{(T_q)}_{x^n}   &=:X_{x^n}  \approx W_{x^n}, \\
    \E X^{(T_q)}_{x^n|k} &=:X_{x^n|k}\approx W_{x_k}.
  \end{align*}
  Denote the minimal nonzero entry of $W$ by $u$, and choose
  $\epsilon$ so small that for all typical $x^n$ and all $k$
  $$X^{(\nu)}_{x^n|k}\geq \frac{u}{2}\ \text{ on }\supp W_{x_k}.$$
  By lemma~\ref{lemma:large:deviation} we obtain
  \begin{equation*}\begin{split}
    \Pr &\left\{\frac{1}{Q}\sum_{q=1}^Q X^{(T_q)}_{x^n|k}
               \not\in[(1\pm\epsilon)X_{x^n|k}]\text{ on }\supp W_{x_k}\right\}      \\
        &\hspace{4.7cm} \leq 2|\fset{Y}|\exp\left(-Q\frac{\epsilon^2 u}{4\ln 2}\right).
  \end{split}\end{equation*}
  Hence the sum of these probabilities is
  upper bounded by
  $$2|\fset{Y}|\,|\fset{X}^n|\exp\left(-Q\frac{\epsilon^2 u}{4\ln 2}\right),$$
  which is less than $1$ for
  $$Q> \frac{4\ln 2}{\epsilon^2 u}\left(n\log|\fset{X}|+\log(2|\fset{Y}|)\right).$$
  Hence there exist actual values $T_1,\ldots,T_Q$ such that
  $$\forall x^n\in\fset{T}^n_{P,\delta} \forall k\quad
      \frac{1}{2}\left\|W_{x_k}-\frac{1}{Q}\sum_q D_{T_q}(E_{T_q}(x^n))_k\right\|_1
                                                                   \leq 3\epsilon,$$
  which is what we wanted to prove: observe that $Q$ grows only polynomially.
\end{beweis}
\par\medskip
As we remarked already in the introduction,~\cite{kramer:savari} proposed to
prove this result (and indeed more, being interested in the tradeoff between
rate and error), but eventually turned to the much softer
condition~(\ref{eq:ks:condition}), which originates from the traditional model of
rate distortion theory.

\section{A general construction}
\label{sec:howards}
Nice though the idea of the previous section is, the lower bound results
show that on this road we cannot hope to approach the conjectured bound,
because without common randomness at hand we have to spend
communication at the same rate to establish it
(compare~\cite{mbcc:sim:ent}, appendix, for this
rather obvious--looking fact).
\par
In this section we want to study the \emph{perfect}
restitution of the probability distributions $W_x$ (i.e.~$\lambda=0$):
\par
Recall that here we want to minimize $H(E(X^n))$, and this minimum we call
$S(n)$. Obviously $S(n_1+n_2)\leq S(n_1)+S(n_2)$, so the limit
$$\Gamma(P,W)=\lim_{n\rightarrow\infty} \frac{1}{n}S(n)$$
exists, and is equal to the infimum of the sequence.
\par
Then we have
\begin{satz}
  \label{satz:old:new}
  For all $\lambda\in(0,1)$
  $$\limsup_{n\rightarrow\infty} \frac{1}{n}\log M(n,\lambda)\leq \Gamma(P,W).$$
\end{satz}
\begin{beweis}
  It is sufficient to prove the inequality for $S(1)$
  in place of $\Gamma(P,W)$:
  \par
  Fix a $1$--code $(e,d)$ with $H(e(X))=S(1)$.
  Then, for $n\geq 1$ choose any $(n,\lambda)$--source
  code $(F,G)$ for $e(X_1)\ldots e(X_n)$, which is possible
  at rate $H(e(X))+o(1)$.
  Then $(E,D)$ with $E=F\circ e^n$ and $D=d^n\circ G$
  is an $(n,\lambda)$--code for the mixed state source
  with limiting rate $S(1)$.
\end{beweis}
It would be nice if we could prove also an inequality in the
other direction, but it seems that a direct reduction like in
the previous proof does not exist: for this we would need to take
an $(n,\lambda)$--code and convert it to an $(n,0)$--code, increasing the
entropy only slightly.
\par
\begin{figure}[ht]
  \label{fig:network}
  \includegraphics{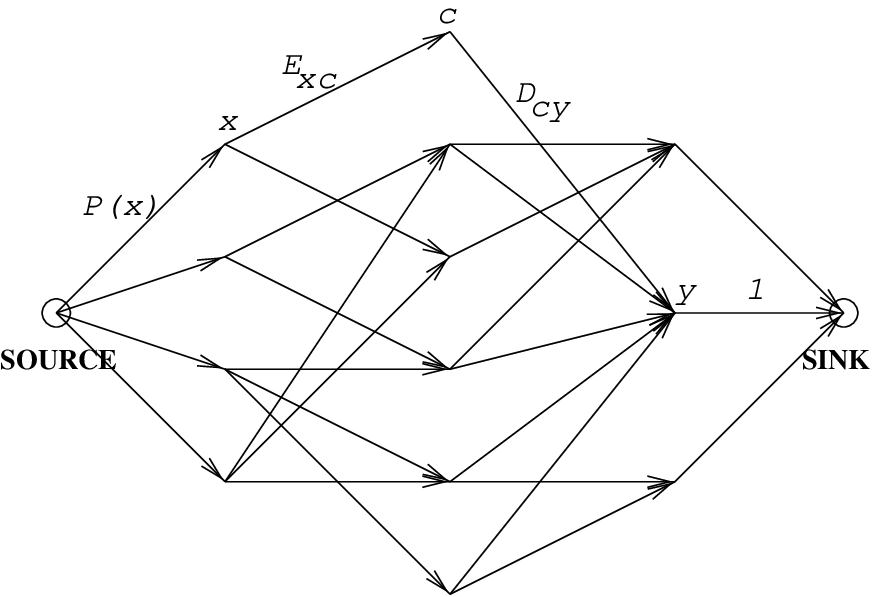}
  \caption{The probability flow network to simulate the distributions $W_x$.
    Note that we included a sink, edges leading to the sink
    obviously having probability $1$.}
\end{figure}
A nice picture to think about the problem of finding $S(1)$
is the following in the spirit of flow networks:
\par
From the source we go to one of the nodes $x\in\fset{X}$,
with probability $P(x)$. Then, with a probability
of $E_{xc}=E(x)(c)$ we go to $c\in\fset{C}$, and from there
with a probability of $D_{cy}=D(c)(y)$ to $y\in\fset{Y}$.
Then the condition is that
$$\forall x\in\fset{X}\forall y\in\fset{Y}\quad
                   W_x(y)=\sum_{c\in\fset{C}} D_{cy}E_{xc}.$$
Examples of this constructions are discussed in~\cite{bcfjs} (where
it was in fact invented), and here we
want to add some general remarks on optimizing it, as well
thoughts on a possible algorithm to do that.
\par
We begin with a general observation on the number of intermediate nodes:
\begin{satz}[``c dice with d sides'']
  \label{satz:cd}
  An optimal zero error code for $W$ requires at most $CD-1$ intermediate
  nodes, with $C=|\fset{X}|$, $D=|\fset{Y}|$.
\end{satz}
\begin{beweis}
  For a fixed set $\fset{C}$ the problem is the following:
  \\
  Under the constraints
  \begin{align}
     \label{eq:cd:E}
    \forall xc\ E_{xc}\geq 0,\quad & \forall x\ \sum_c E_{xc}=1,\\
     \label{eq:cd:D}
    \forall cy\ D_{cy}\geq 0,\quad & \forall c\ \sum_y D_{cy}=1,\\
     \label{eq:cd:ED}
                                   & \forall xy\ \sum_c E_{xc}D_{cy}=W_x(y),
  \end{align}
  minimize the entropy $H(\mu)$, where
  $\mu_c=\sum_x P_x E_{xc}$.
  \par
  Observe that for each fixed set of $D_{cy}$ the constraints define
  a convex admissible region for the $E_{xc}$, of which a \emph{concave}
  function is to be minimized. Hence, the minimum will be achieved
  at an extreme point of the region, that we rewrite as
  follows:
  \begin{equation*}\begin{split}
    & \left\{E_{xc}\geq 0:\forall xy\ \sum_c E_{xc}D_{cy}=W_x(y)\right\} \\
    &\phantom{====}
      =\bigoplus_x \left\{E_{xc}\geq 0:\forall y \sum_c E_{xc}D_{cy}=W_x(y) \right\}.
  \end{split}\end{equation*}
  An extreme point must be extremal in every of the summand
  convex bodies $B_x$.
  On the other hand, an extreme point of $B_x$ must meet $\dim B_x$
  many  of the inequalities ($E_{xc}\geq 0$) with equality.
  Since $\dim B_x\geq |\fset{C}|-D$ there remain only
  at most $D$ nonzero $E_{xc}$ for every $x$.
  In particular, only at most $CD$ many $c\in\fset{C}$ are
  accessed at all. In fact, to minimize $H(\mu)$,
  at most $CD-1$, otherwise $c$
  would contain full information about $x$.
\end{beweis}
\begin{rem}
  \label{rem:cd:more}
  The last argument can be improved: for $C,D\geq 2$ we can
  even assume $|\fset{C}|\leq CD-C+1$.
\end{rem}
\par
The argument of the proof gives us the idea that maybe by an alternating
minimization we can find the optimal code:
\par
Indeed, conditions~(\ref{eq:cd:E}) and~(\ref{eq:cd:ED}) for fixed $D$ are
linear in $E$, and the target function is concave (entropy of a linear function
of $E$), so we can find it's minimum at an extreme point of the admissible region.
This part is solved by standard convex optimization methods.
On the other hand, for fixed $E$, eqs.~(\ref{eq:cd:D}) and~(\ref{eq:cd:ED})
are linear in $D$. However, variation does not change the aim function.
Still we have freedom to choose, and this might be
a good rule: let $D$ \emph{maximize} the conditional entropy
$H(D|\mu)$. The rationale is that this entropy signifies the
ignorance of the sender about the actual output. If it does not approach
$H(W|P)$ in the limit this means that the protocol simulates partial feedback
of the channel $W$, which could be used to extract common randomness.
This amount is a lower bound to what the protocol has to communicate in excess
of $I(P;W)$. We have, however, no proof that this rule converges to an optimum.

\section{Applications}
\label{sec:applications}
In this section we point out three important connections to other
questions, some of which depend on positive answers to the
questions~\ref{quest:main} and~\ref{quest:main}'.

\subsection{Common randomness}
It is known that if two parties (say, Alice and Bob) have access
to many inpendent copies of the pair of random variables $(X,Y)$
(which are supposed to be correlated), then they can, by public
discussion (which is overheard by an eavesdropper), create common
randomness at rate $I(X\wedge Y)$, almost independent of the
eavesdropper's information. For details see~\cite{ahlswede:csiszar:1},
where this is proved, and also the optimality of the rate.
One might turn around the question and ask, how much common randomness
is required to create the pair $(X^n,Y^n)$ approximately. This
question, in the vein of that of the previous subsection, is really
about reversibility of transformations between different appearances
of correlation. Note that this was confirmed in~\cite{collins:popescu}
for the case of deterministic correlation between $X$ and $Y$,
i.e.~$H(Y|X)=H(X|Y)=0$, which there was parallelled to entanglement
concentration and dilution for pure states.
\par
An affirmative answer to question~\ref{quest:main}, surprisingly implies
that a rate of $I(X\wedge Y)$ of common randomness is sufficient,
\emph{with no further public discussion} to create pairs $X,Y$.
This is done by first creating the distribution $Q$ of
$E(X^n)$ on $\fset{C}$ from the common randomness (this Alice and Bob
do each on their own!):
this may be not altogether obvious as the common randomness is
assumed in pure form (i.e. a uniform distribution on $N$ alternatives),
while the distribution $Q$ may have no regularity.
To overcome this difficulty fix an $\epsilon>0$ and let
$$k=\left\lceil\frac{\log|\fset{C}|-\log\epsilon}{\epsilon}\right\rceil.$$
Now we partition the unit interval into the subintervals
\begin{align*}
  I_a      &=\left[(1+\epsilon)^{-(a-1)},(1+\epsilon)^{-a}\right),\ a=1,\ldots,k, \\
  I_\infty &=[(1+\epsilon)^{-k},0],
\end{align*}
and define $\fset{C}_a=\{c\in\fset{C}:Q(c)\in I_a\}$, $q_a=Q(\fset{C}_a)$.
Notice that for $a<\infty$ the probabilities for $c$'s belonging to the
same set $\fset{C}_a$ differ
from each other only by a factor between $1-\epsilon$ and $1+\epsilon$, and that
$q_\infty\leq \epsilon$, because of $(1+\epsilon)^{-k}\leq \epsilon/|\fset{C}|$,
by definition of $k$.
Hence, defining uniform distributions $U_a$ on $\fset{C}_a$ for $a<\infty$,
it is immediate that
$$\frac{1}{2}\left\|Q-\sum_{a=1}^k \frac{q_a}{1-q_\infty} U_a\right\|_1
                                                            \leq 2\epsilon.$$
Now the distribution on the $a=1,\ldots,k$ in this formula can be approximated
to within $1/k$ by a $k^2$--type distribution, which in turn can be obtained
directly from a uniform distribution on $k^2$ alternatives.
In this way we reduced everything to a number of uniform distributions,
maybe on differently sized sets, all bounded by $|\fset{C}|$ and
a helper uniform distribution on a set of size $k^2$.
However, it is well known that these can be obtained from a uniform
distribution on $O(k^2|\fset{C}|)$ items within arbitrarily small error.
\par
Given this distribution on $\fset{C}$, Bob applies $D$, whereas Alice applies the transpose
channel $E'$ to $E$. One readily checks that this produces the joint
distribution of $X^n,Y^n$, up to arbitrarily small disturbance in the
total variational norm.
\par
Note that this result would imply a new proof of the optimality of
of the rate $I(X \wedge Y)$ of common randomness distillation
from $X^n,Y^n$: because we can simulate the latter pair of random variables
with this rate of common randomness, we would obtain a net increase
of common randomness after application of the distillation, which clearly
cannot be.

\subsection{Channel coding}
\label{subsec:channel:coding}
It was already pointed out that this study has the paper~\cite{bennett:et:al}
as one motiviation, with its idea to prove the optimality of Shannon's
coding theorem by showing that
every noisy channel $W$ can be simulated by
a binary noiseless one operating at rate $C(W)$. Shannon's theorem
is understood as saying that the noisy channel can simulate a binary noiseless
one of rate $C(W)$. Both simulations are allowed to perform with small
error. Note that an affirmation of question~\ref{quest:main}',
implies that this can be done, without the common randomness
consumption like in section~\ref{sec:cr:trick}.
As indicated, this provides a proof of the converse to Shannon's
coding theorem:
\par
The idea is that otherwise we could, given a rate of $C(W)$ noiseless
bits simulate the channel, which in turn could be used to transmit
at a rate $R>C(W)$. The combination of simulation and coding yields
a coding method for transmitting $R$ bits over a channel providing
$C(W)$ noisless bits, which is absurd (in~\cite{bennett:et:al} this
reasoning is called ``causality argument'').
Theorem~\ref{satz:sim:feedback:channel} allows us to prove even more:
\begin{satz}[Shannon~\cite{shannon}]
  \label{satz:feedback:capacity}
  For the channel $W$ with noisless feedback (i.e. after each symbol $x$
  transmitted the sender gets a copy of the symbol $y$ read by the receiver, and
  may react in her encoding)
  the capacity is given by $C(W)$. In fact, for the maximum size
  $M_{\rm f}(n,\lambda)$ of an $(n,\lambda)$--feedback code
  $$M_{\rm f}(n,\lambda)\leq \exp\bigl(nC(W)+O(\sqrt{n})\bigr).$$
\end{satz}
\begin{beweis}
  Let an optimal $(n,\lambda)$--feedback code for the channel $W^n$ with
  noiseless feedback be given.
  We will construct an $(n^2,\lambda')$--code with shared randomness,
  as follows: 
  \par
  Choose a simulation of the channel $W$ on $n$--blocks sending
  $nC(W)+O(\sqrt{n}\log n)$ bits, and using shared randomness, and with
  error bounded by $\epsilon=\frac{1-\lambda}{2n}$
  (this is possible by the construction of theorem~\ref{satz:sim:feedback:channel}
  --- see remark~\ref{rem:chernoff:better}).
  We shall use $n$ independent copies of the feedback code in
  parallel: in each round $n$ inputs symbols are prepared,
  sent through the channel, yielding $n$ respective feedback
  symbols. Obviously, each round can be simulated with
  an error in the output distribution bounded by $\epsilon$,
  using our simulation of the channel W (which, as we remarked
  earlier, simulates even the feedback).
  In each of the parallel executions of the feedback code
  thus accumulates an error of at most $\frac{1-\lambda}{2}$,
  increasing the error probability of the code to $\frac{1+\lambda}{2}$.
  Hence on the block of all the $n$ feedback codes we can bound
  the error probability by $\lambda'=1-\left(\frac{1-\lambda}{2}\right)^n$.
  \par
  But this is subexponentially (in $N=n^2$) close to $1$, so
  a standard argument applies:
  \par
  First, by considering average error probability we can get
  rid of the shared randomness: there exists one value of the
  shared random variable for which the average error probability
  is bounded by $\lambda'$.
  Then we can argue that there is a subset $\fset{U}$ of the
  constructed code's message set $\fset{M}^n$ which
  has \emph{maximal} error probability bounded by
  $\lambda''=\frac{1+\lambda'}{2}$ and
  $$|\fset{U}|\geq (1-{\lambda}'')|\fset{M}^n|.$$
  \par
  What we achieved so far hence is this: a code of $|\fset{U}|$
  messages with error probability $\lambda''$ and using
  $NC(W)+o(N)$ noiseless bits. Clearly, we may assume the encoder
  to be deterministic without losing in error probability.
  But then at most $(1-{\lambda}'')^{-1}$ messages can be mapped
  to the same codeword without violating the error condition.
  \par
  Collecting everything we conclude
  \begin{equation*}\begin{split}
    |\fset{M}|^n &\leq (1-{\lambda}'')^{-1} |\fset{U}|                      \\
                 &\leq (1-{\lambda}'')^{-2} \exp\left(n^2C(W)+o(n^2)\right) \\
                 &=    \left[\exp\bigl(nC(W)+o(n)\bigr)\right]^n,
  \end{split}\end{equation*}
  implying the theorem.
\end{beweis}
\begin{rem}
  \label{rem:feedback}
  The \emph{weak converse} (i.e.~the statement that the rate for codes with
  error probability approaching $0$ is bounded by $C(W)$) is much easier
  to obtain, by simply keeping track of the mutual information between
  the message and the channel output through the course of operating a
  feedback code, using some well--known information identities, and finally
  estimating the code rate employing Fano's inequality.
\end{rem}

\subsection{Rate--distortion theorem}
\label{subsec:rate:dist}
Let $d:\fset{X}\times\fset{Y}\longrightarrow \R_{\geq 0}$ be any
\emph{distortion measure}, i.e. a non--negative real function.
This function is extended to words $\fset{X}^n\times\fset{Y}^n$
by letting
$$d^n(x^n,y^n)=\sum_{k=1}^n d(x_k,y_k).$$
Shannon's rate distortion theorem is about the following
problem: construct an $n$--block code $(E,D)$ (which my be chosen
to be deterministic) such that for a given $d\geq 0$
$$d(E,D):=\sum_{x^n} P^n(x^n) d^n(x^n,D(E(x^n))) \leq nd,$$
i.e., the average distortion between source and output word
is bounded by $nd$.
\par
A pair $(R,d)$ of non--negative real numbers is said to be
\emph{achievable} if there exist $n$--block codes with code
rate tending to $R$ and distortion rate asymptotically
bounded by $d$.
Define the \emph{rate--distortion function} $R(d)$ as
the minimum $R$ such that $(R,d)$ is achievable.
\begin{satz}[Shannon~\cite{shannon:rd}]
  \label{satz:shannon:r:d}
  The rate distortion function is given by the following formula:
  $$R(d)=\min \left\{I(P;W):\ W\text{ channel s.t. }\E d(X,Y)\leq d\right\},$$
  where $\E d(X,Y)=\sum_{xy} P(x)W_x(y)d(x,y)$ is the expected
  (single--letter) distortion when using the channel $W$.
\end{satz}
The proof of ``$\geq$'' here is a simple exercise using convexity
of mutual information in the channel and standard entropy inequalities.
We can give a simple proof of the ``$\leq$''--part of this result, using
theorem~\ref{satz:feedback:channel:opt}:
\par
Choose some channel $W$ satisfying the
distortion constraint. Then mapping $x^n$ to $W^n_{x^n}$ obviously
satisfies the distortion constraint on the code in the sense that the expected
distortion between input and output, over source and channel, is bounded by $nd$.
Of course, sampling $W^n_{x^n}$ at the encoder and sending some $y^n$ will
not meet the bound $I(P;W)$. However, we can apply theorem~\ref{satz:feedback:channel:opt}
to approximately simulate the joint distribution of $x^n$ and $y^n$ by using some common
randomness $\nu$ and a deterministic code $(E_\nu,D_\nu)$ sending
$nI(P;W)+O(\sqrt{n})$ bits. Hence, invoking linearity of the definition
of $d(E,D)$,
$$\sum_\nu x_\nu d(E_\nu,D_\nu)\leq nd+O(\epsilon),$$
so there must be one $\nu$ such that $d(E_\nu,D_\nu)\leq nd+O(\epsilon)$,
which ends our proof.
\par\medskip
At this point we would like to advertise our point of view that
theorem~\ref{satz:coding:local}, and even more so
theorem~\ref{satz:sim:feedback:channel}, is what rate--distortion is actually
about: the former theorem shows how to simulate a given channel on all
individual positions of a transmission, and this is what we need in
rate--distortion. In fact, rate--distortion theory is unchanged when instead
of the one convex condition (``distortion bound'') on the code
we have several, effectively restricting the admissible approximate
joint types of input and output to any prescribed convex set --- in particular
a single point.
\par
The strength of theorem~\ref{satz:coding:local} in comparison to such a development
of rate--distortion theory lies in the fact that with its help we satisfy the
convex conditions in \emph{every letter}, not just in the block average.
And theorem~\ref{satz:sim:feedback:channel} gives the analogue of this
even with the condition imposed on the whole block, yielding results
that are not obtainable by simply applying rate--distortion tools
(see e.g.~\cite{q-c-tradeoff}).

\section{Compression of sources of quantum states}
\label{sec:quantum}
The problem studied in this paper has a natural extension to
quantum information theory: now the source emits (generally mixed)
quantum states $W_x$ on the Hilbert space ${\cal Y}$
($x\in\fset{X}$), with probabilities $P(x)$,
and an $(n,\lambda)$--code is a pair
$(E,D)$ of maps
\begin{equation}
  \label{eq:qu:code}
  \begin{array}{c}
    E:\fset{X}^n        \longrightarrow \alg{S}({\cal C}),  \\
    D:\alg{S}({\cal C}) \longrightarrow \alg{S}({\cal Y}^{\otimes n}),
  \end{array}
\end{equation}
where $\alg{S}({\cal C})$ is the set of states
on the code Hilbert space ${\cal C}$ and $D$ is
completely positive, trace preserving, and linear.
The condition to satisfy is
\begin{equation}
  \label{eq:qu:condition}
  \sum_{x^n\in\fset{X}^n} P^n(x^n)\frac{1}{2}\|W^n_{x^n}-D(E(x^n))\|_1\leq \lambda,
\end{equation}
with the trace norm $\|\cdot\|_1$ on density operators.
Define, like before, $M(n,\lambda)$ as the minimum $\dim{\cal C}$
of an $(n,\lambda)$--code.
\par
Sometimes, the stronger condition
\begin{equation}
  \label{eq:qu:condition:strong}
  \forall x^n\in\fset{T}^n_{P.\delta}\quad \frac{1}{2}\|W^n_{x^n}-D(E(x^n))\|_1\leq \lambda
\end{equation}
will be applied.
\par
Notice that this contains our original problem as the special case
of a \emph{quasiclassical} ensemble, when all the $\rho_x$ commute
(which means they can be interpreted as probability distributions
on a set of common eigenstates).
\par
This problem (with a number of variations, which we explained
in the introductory section~\ref{sec:pdsources} for the classical
case) is studied in~\cite{bcfjs}.
There (and previously in~\cite{horodecki}) it is shown that the
lower bound theorem~\ref{satz:lower} holds in the quantum case,
too, with understanding $H$ as von Neumann entropy:
\begin{satz}
  \label{satz:mixed:lower}
  For all $n$, $\lambda$
  $$\frac{1}{n}M(n,\lambda)\geq I(P;W)-f(\lambda),$$
  with a function $f(\lambda)\rightarrow 0$ for $\lambda\rightarrow 0$.
  \qed
\end{satz}
Let us improve this slightly by proving the strong version of this
result:
\begin{satz}
  \label{satz:mixed:strong:lower}
  For all $\lambda\in(0,1)$
  $$\liminf_{n\rightarrow\infty}\frac{1}{n}M(n,\lambda) \geq I(P;W).$$
\end{satz}
\begin{beweis}
  By much the same method as the proof of theorem~\ref{satz:strong:lower}:
  the changes are that we need the more general code selection
  result of~\cite{winter:diss}, thm.~II.4, instead of the classical
  theorem~\cite{csiszar:koerner}, and which we state separately below:
  if $(E,D)$ is an optimal $(n,\lambda)$--code, define
  $$\fset{A}=\left\{x^n:\ \frac{1}{2}\|W^n_{x^n}-D(E(x^n))\|_1\leq \sqrt{\lambda}\right\}.$$
  Obviously $P^n(\fset{A})\geq 1-\sqrt{\lambda}>0$, so we can apply
  lemma~\ref{lemma:max:code} and find an $(n,\epsilon)$--transmission
  code $\fset{U}\subset\fset{A}$ for $W^n$ such that
  $$|\fset{U}|\geq \exp\bigl(nI(P;W)-O(\sqrt{n})\bigr).$$
  This is an $(n,\lambda')$--code for the channel $D\circ E$, with
  $\lambda'=\sqrt{\lambda}+\epsilon<1$, if we choose $\epsilon$ small enough.
  Combining $E$ with the transmission encoder, and $D$ with the
  transmission decoder, we obtain an $(n,\lambda)$--transmission
  code for $|\fset{U}|$ many messages over a noiseless system with Hilbert
  space ${\cal C}$ of dimension $M(n,\lambda)$.
  \par
  To each message $u\in\fset{U}$ there belongs a decoding operator $\Delta_u\geq 0$
  on the coding space ${\cal C}$, forming together a POVM: $\sum_u \Delta_u=\1$.
  Now to decode correctly with probability $1-\lambda'$, for each $u$ we must have
  $$\tr\Delta_u\geq 1-\lambda'.$$
  On the other hand, by $\sum_u \tr\Delta_u=M(n,\lambda)$, we conclude
  \begin{equation*}\begin{split}
    M(n,\lambda) &=    \dim{\cal C} \geq \frac{1}{1-\lambda'}|\fset{U}|        \\
                 &\geq \frac{1}{1-\lambda'}\exp\bigl(nI(P;W)-O(\sqrt{n})\bigr),
  \end{split}\end{equation*}
  and we are done.
\end{beweis}
\begin{lemma}
  \label{lemma:max:code}
  For $0<\tau,\lambda<1$ there is a constant $K'$ and $\delta>0$ such that
  for every discrete memoryless quantum channel $W$ and distributions $P$ on 
  $\fset{X}$ the following holds: if
  $\fset{A}\subset\alg{W}^n$ is such that $P^n(\fset{A})\geq\tau$
  then there exists an $(n,\lambda)$--transmission code
  $(E,D)$ with the properties
  $$\forall m\in\fset{M}\quad
               E(m)\in\fset{A}\text{ and }\tr D_m\leq\tr\Pi^n_{H,f(m),\delta}\ ,$$
  $$|\fset{M}| \geq \exp\bigl(nI(P;W)-K'\sqrt{n}\bigr).$$
\end{lemma}
\begin{beweis}
  See~\cite{winter:diss}, thm.~II.4.
\end{beweis}
\par
Progress on the problem of achievability of this bound is not
known to us. It is remarkable that Koashi and Imoto~\cite{koashi:imoto}
could obtain the exact optimal bound in the case of \emph{blind}
coding. It is indirectly defined via a canonical joint decomposition of the
source states, but it can be derived from their result that generically
the optimum rate is $H(PW)$, which is achieved by simply Schumacher
encoding the ensemble $\{P(x),W_x\}$.
\par
Nevertheless, the results obtained in the classical case are very
encouraging, so we state two conjectures:
\begin{conj}
  \label{conj:qmain:cr}
  For $0<\lambda<1$ there exist $(n,\lambda)$--codes with common randomness,
  asymptotically achieving transmission rate $I(P;W)$ and common
  randomness consumption $H(W|P)$.
\end{conj}
\par
If it turns out true, and also question~\ref{quest:main} has a positive answer,
we might even hope that also 
\begin{quaest}
  \label{quest:qmain}
  For $0<\lambda<1$, is
  $$\lim_{n\rightarrow\infty} \frac{1}{n}\log M(n,\lambda)=I(P;W)\ ?$$
  [Note that, as in the case of question~\ref{quest:main}, codes achieving the optimal
  bound may also be constructed to satisfy eq.~(\ref{eq:qu:condition:strong}).]
\end{quaest}
answers ``yes''.
\par
The implications of these statements, if they are true, would be of great
significance to quantum information theory: not only would we get
a new proof of the capacity of a classical--quantum channel being
bounded by the maximum of the Holevo information and for the 
optimality of common randomness extraction from a class of bipartite
quantum sources~\cite{wilmink:thesis}, but also the achievability of
$I(P;W)$ in the quantum rate distortion problem~\cite{q:rate:d}
with visible coding would follow, that until now has escaped all attempts.

\section{Concluding remarks}
\label{sec:conclusion}
We demonstrated the current state of knowledge in the problem of
visible compression of sources of probability distributions and
its extension to mixed state sources in quantum information theory.
Apart from reviewing the currently known constructions we
contributed a better understanding of the resources involved:
in particular the use of common randomness in some of them,
and providing strong converses. Also we showed the numerous applications
the result (and sometimes the conjectures) have throughout information
theory, making the matter an eminent unifying building block
within the theory.
\par
We would like to draw the attention of the reader once more
to our questions~\ref{quest:main} and~\ref{quest:qmain},
and especially the conjecture~\ref{conj:qmain:cr}
offering them as a challenge to continue this work.


\section*{Acknowledgments}
Research partially supported by SFB 343 ``Diskrete Strukturen in
der Mathematik'' of the Deutsche Forschungsgemeinschaft, by Fakult\"at f\"ur
Mathematik, Universit\"at Bielefeld, by the University of Bristol, and by
the U.K. Engineering and Physical Sciences Research Council.
\par
I would like to thank Richard Jozsa for numerous conversations on mixed state
compression, in particular on the content of section~V.


\end{document}